\documentclass[twocolumn,aps,prl,amsmath,floatfix]{revtex4}
\usepackage{graphicx}
\begin{document}

\title{Effect of Dynamical Coulomb Correlations on the Fermi Surface 
       of  Na$_{0.3}$CoO$_2$}
\author{ H. Ishida$^1$, M.D. Johannes$^2$, and A. Liebsch$^3$}
\affiliation{
\mbox{
$^1$College of Humanities and Sciences, Nihon University,~Sakura-josui,
        ~Tokyo 156, Japan}  \\
\mbox{
$^2$Code 6391, Naval Research Laboratory, Washington, D.C.  20375, 
    USA} \\
\mbox{
$^3$Institut f\"ur Festk\"orperforschung,~Forschungszentrum J\"ulich,
        ~52425 J\"ulich, Germany} \\
}

\begin{abstract}
The $t_{2g}$ quasi-particle spectra of Na$_{0.3}$CoO$_2$ are calculated 
within the dynamical mean field theory. It is shown that as a result 
of dynamical Coulomb correlations charge is transfered from the nearly 
filled $e_{g'}$ subbands to the $a_{1g}$ band, thereby reducing orbital 
polarization among Co $t_{2g}$ states.  Dynamical correlations 
therefore stabilize the small $e_{g'}$ Fermi surface pockets, in contrast 
to angle-resolved photoemission data, which do not reveal these pockets.\\
\end{abstract}
\maketitle

The intercalated layer compound Na$_x$CoO$_2$ has been intensely studied 
during the recent years since as a function of Na doping it exhibits a 
variety of fascinating properties. The relevant valence bands near the 
Fermi level consist of Co $t_{2g}$ states with occupancy
$3d^{5+x}$. For $x\approx 0.50-0.75$ an unusually large thermopower is 
observed \cite{terasaki}. For $x\approx0.3$ hydration  
gives rise to a superconducting transition at 4.5~K \cite{takada}. 
In a narrow region near $x=0.5$ the material undergoes a metal 
insulator transition \cite{foo,wang}. The end-member at $x=0$ with 
a single hole per Co atom is believed to be a Mott insulator.
At $x=1$, since the filled $t_{2g}$ bands are separated by about 
1.5~eV from the empty $e_g$ bands, one finds a band insulator.

Despite considerable experimental and theoretical effort,
fundamental electronic properties of  Na$_x$CoO$_2$ such as the 
qualitative topology of the Fermi surface remain controversial and
are not well understood.    
In the metallic phase the hexagonal Fermi surface predicted by band
theory within the local density approximation (LDA) consists of a large
hole pocket centered at $\Gamma$ and six small hole pockets near $K$ 
\cite{singh} (see Fig.~1). Because of the layered structure the $t_{2g}$ 
levels split into a $a_{1g}$ level and doubly degenerate $e_{g'}$ levels.      
The large Fermi surface stems mainly from the $a_{1g}$ bands while the 
small hole pockets have predominantly $e_{g'}$ character. The existence 
of these hole pockets is thought to be crucial for the understanding of 
the superconducting phase \cite{super}. On the other hand, angle resolved 
photoemission spectra (ARPES) for $x=0.7$ \cite{hasan1} and $x=0.6$ 
\cite{yang1} provide evidence only for $a_{1g}$ bands crossing the Fermi 
level. These data suggest that the $e_{g'}$ bands are filled
due to inter-orbital charge transfer not described within the LDA. New ARPES
data for $x=0.3\cdots0.7$ \cite{yang2} and $x=0.3$ \cite{hasan2} 
also do not show the small $e_{g'}$ hole pockets.

To address the question of possible modifications of the Fermi surface 
via Coulomb correlations not included in the LDA, several theoretical 
studies were carried out for the half-metallic, magnetic region near
$x=0.7$ \cite{ldau}, using the LDA+U approach \cite{LDAU}. For a relatively 
large local Coulomb energy ($U>3$~eV), the $e_{g'}$ subbands are indeed
filled. 
In the metallic phase, however, in particular, for Coulomb energies comparable
to the single-electron band width, it is well known that correlations have an 
important dynamical component not captured within the LDA+U.      

\begin{figure}[b!]
  \begin{center}
  \vskip-6mm
  \includegraphics[width=5.5cm,height=5cm,angle=  0]{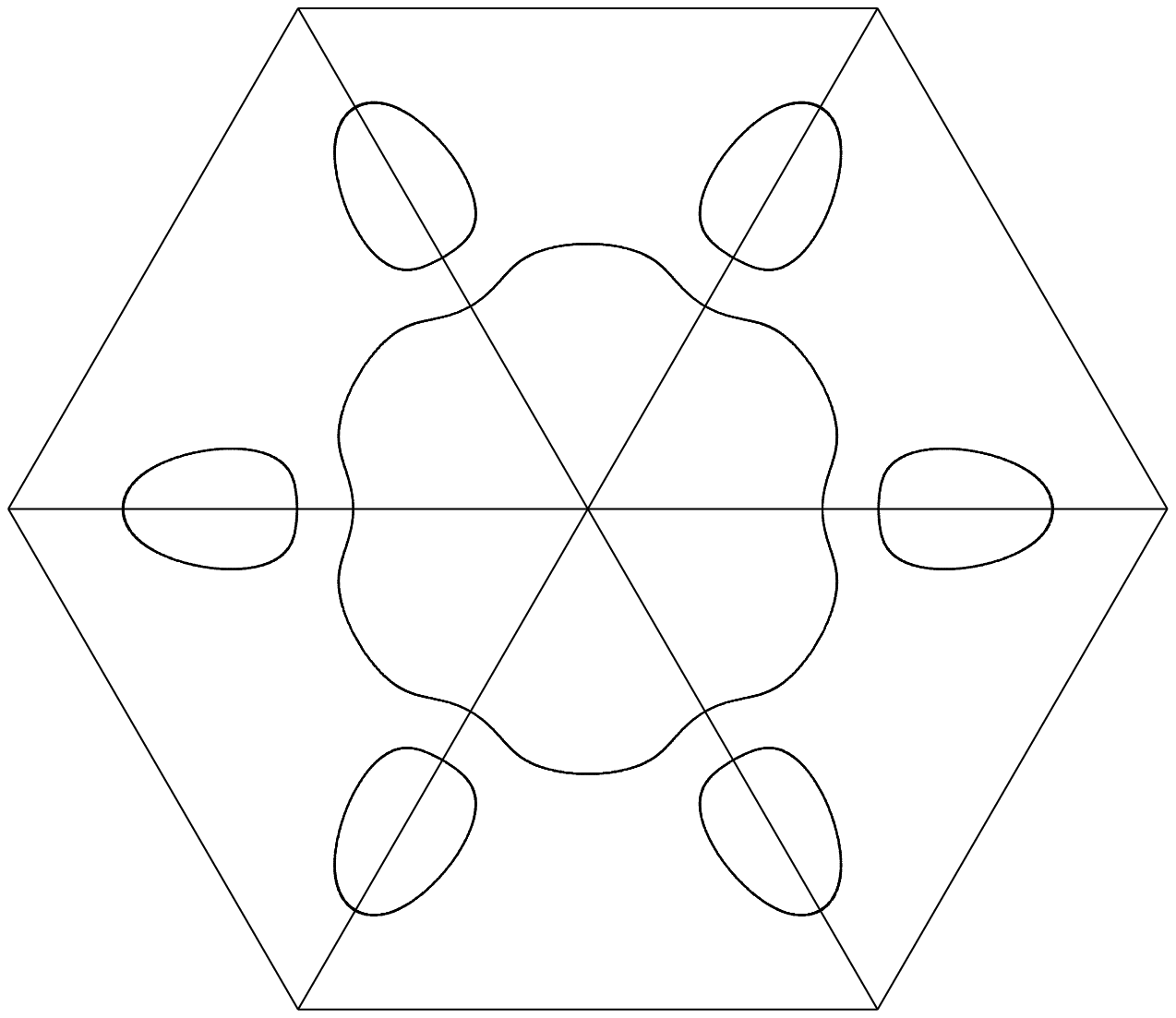}
  \vskip-6mm
  \includegraphics[width=4.5cm,height=7cm,angle=-90]{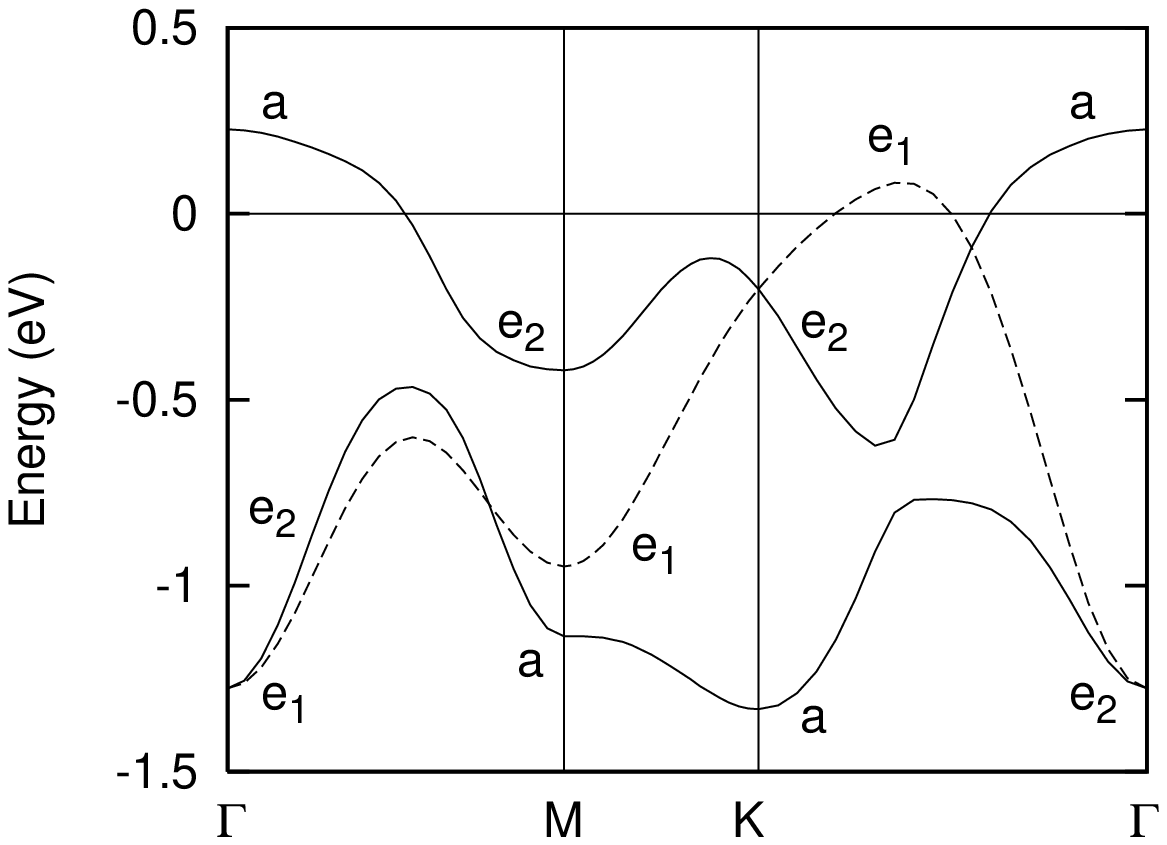}
  \end{center}
  \vskip-4mm
\caption{
Upper panel: Fermi surface of Na$_{0.3}$CoO$_2$. 
Lower panel: tight-binding fit to LDA bands; $E_F=0$. 
Solid curves: even bands; dashed curve: odd band. 
The $e_{g'}$ states above $E_F$ give rise 
to the small hole pockets of the Fermi surface.
}\end{figure}

The aim of this work is to elucidate the possibility of modifying the 
Fermi surface of Na$_{0.3}$CoO$_2$ via dynamical Coulomb correlations.
More specifically we focus on the charge transfer between the paramagnetic
$t_{2g}$ subbands. Since only the total electron number is conserved when
correlations are taken into account, the occupations of individual
subbands may vary with the strength of the local intra- and inter-orbital 
Coulomb energies. In a single-band picture, the key effect of 
dynamical fluctuations is the spectral weight transfer from the 
quasi-particle peak near $E_F$ to the incoherent satellites associated 
with the lower and upper Hubbard bands. In a multi-band material, this
spectral weight transfer is orbital dependent, opening the possibility 
of redistributing electronic charge among the valence orbitals and 
modifying the shape of the Fermi surface. 

To investigate these multi-band correlation effects we use the 
dynamical mean field theory (DMFT) combined with the Quantum Monte Carlo 
(QMC) method \cite{DMFT}. The remarkable result of this work is that in 
the metallic domain of Na$_x$CoO$_2$ near $x=0.3$ dynamical correlations 
shift charge from the $e_{g'}$ states to the $a_{1g}$ bands, thereby 
stabilizing the $e_{g'}$ hole pockets and slightly reducing the $a_{1g}$
Fermi surface. The overall topology of the Fermi surface remains the same
as in the LDA.  
 
Fig.~1 shows a tight-binding fit to the Na$_{0.3}$CoO$_2$ bands 
calculated within the LDA and using the linearized augmented plane wave 
(LAPW) method. We consider the paramagnetic phase observed experimentally 
at $x=0.3$. Since we are interested in the qualitative issue of charge 
transfer between $t_{2g}$ subbands the weak dispersion along the $c$-axis 
is neglected. The full band structure involving Co 3d and O 2p states 
is down-folded to a $3\times3$ Co $t_{2g}$ tight-binding Hamiltonian in 
which on-site energies and hopping integrals represent effective energies 
accounting for direct Co-Co and indirect Co-O-Co interactions. Including 
three neighbor shells, with $dd\sigma$, $dd\pi$ and $dd\delta$ matrix
elements, an excellent fit to the LAPW band structure is achieved. 
The details of the tight binding model and down-folding 
procedure will be given elsewhere \cite{downfolding}.

Because of the planar structure of the system, it is convenient to 
transform the $d_{xy,xz,yz}$ orbitals into $a_{1g}$ and $e_{g'}$ states, 
where  $a_{1g} =  (  d_{xy}+d_{xz}+d_{yz} )/\sqrt3 $,  
     \,$e_{g1} =  (         d_{xz}-d_{yz} )/\sqrt2 $,   
     \,$e_{g2} =  ( 2d_{xy}-d_{xz}-d_{yz} )/\sqrt6 $. 
Quantities such as the local density of states and local quasi-particle 
self-energy are diagonal in this representation. For instance, the 
$a_{1g}$ and $e_{g'}$ density of states are  
$\rho_a= \rho_{ii}+2\rho_{ij}$, $\rho_e= \rho_{ii}-\rho_{ij}$, 
where $\rho_{ii}$ ($\rho_{11}=\rho_{22}=\rho_{33}$)
and $\rho_{i\ne j}$ ($\rho_{12}=\rho_{13}=\rho_{23}$) 
are the diagonal and off-diagonal elements of the $t_{2g}$ density 
of states matrix.

\begin{figure}[t!]
  \begin{center}
  \includegraphics[width=4.5cm,height=7cm,angle=-90]{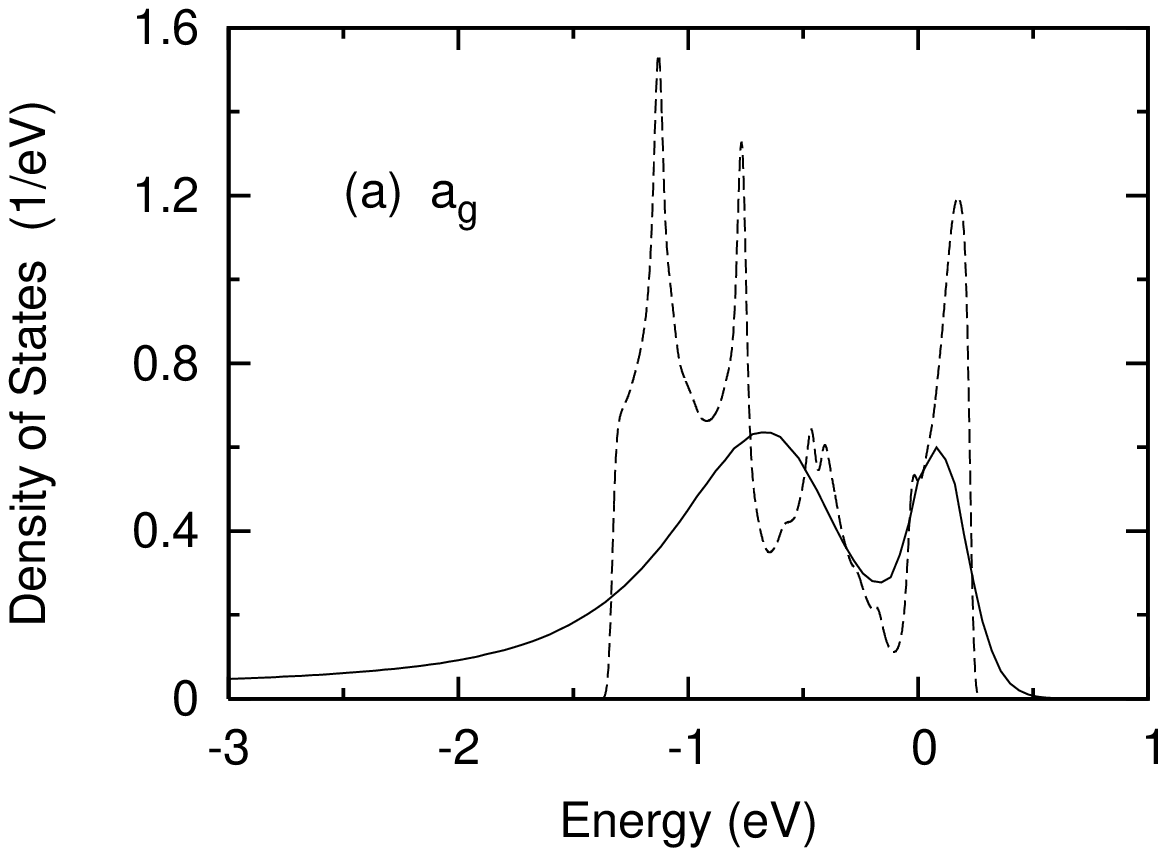}
  \includegraphics[width=4.8cm,height=7cm,angle=-90]{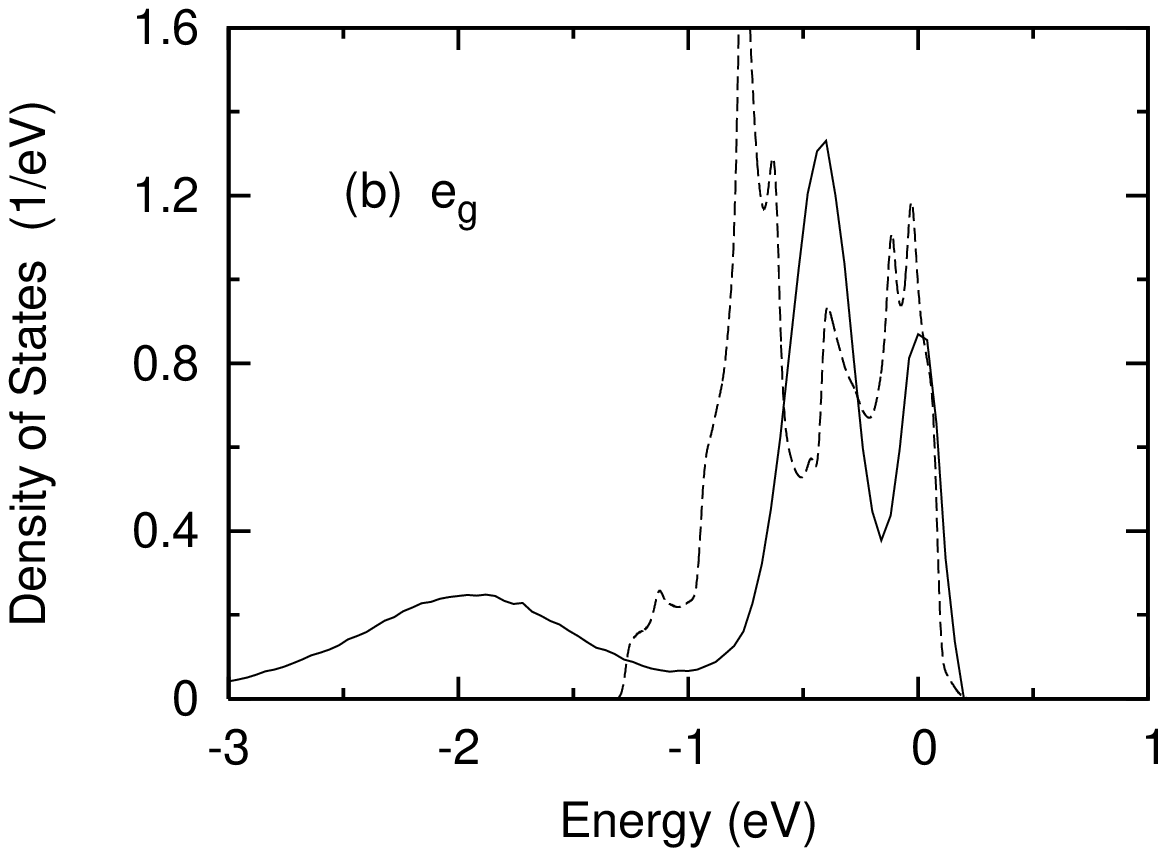}
  \end{center}
  \vskip-4mm
\caption{
Quasi-particle spectra for metallic Na$_{0.3}$CoO$_2$ calculated 
within the DMFT (solid curves) for $U=3$~eV, $J=0.8$~eV; $T=770$~K,
$E_F=0$. Dashed curves: LDA local density of states. 
(a) $a_{1g}$ states; (b) $e_{g'}$ states. 
}\end{figure}

Fig.~2 shows the $a_{1g}$ and $e_{g'}$ quasi-particle spectra for 
Na$_{0.3}$CoO$_2$ as calculated within the DMFT. The local Coulomb 
interaction defining the quantum impurity problem is characterized 
by intra- and inter-orbital matrix elements $U=U_{ii}$ and 
$U'=U_{i\ne j}=U-2J$, where $J$ is the Hund's rule exchange integral. 
The value of $U$ for the entire $t_{2g}$, $e_{g}$ manifold (total width
about 4~eV) was estimated at 3.7~eV \cite{johannes}. For the  
narrower $t_{2g}$ bands a smaller value should be more appropriate to 
account for screening involving empty $e_{g}$ states. Since accurate 
values of $U$ and $J$ are not available, DMFT calculations for several 
values were carried out to study their effect on the inter-orbital 
charge transfer. The temperatures were $T\approx385$ and 
770~K corresponding to about 30 and 60~meV thermal broadening. Up to
$10^6$ sweeps were done in the QMC calculations. The quasi-particle 
spectra are obtained via maximum entropy reconstruction \cite{jarrell}.

The quasi-particle spectra show the characteristic band narrowing near 
$E_F$ caused by dynamical correlations and the transfer of weight from 
the coherent to the incoherent spectral region. In the slightly narrower 
$e_{g'}$ band correlations are strong enough to give rise to a lower 
Hubbard band. Also noticeable is the substantial lifetime broadening of 
valence states due to creation of electron hole pairs. The occupations 
of these distributions are:  
$n_a = 0.853$, $n_e = 0.899$, which should be compared to the LDA values  
$n_a = 0.797$, $n_e = 0.927$. 

\begin{figure}[t!]
  \begin{center}
  \includegraphics[width=4.8cm,height=7cm,angle=-90]{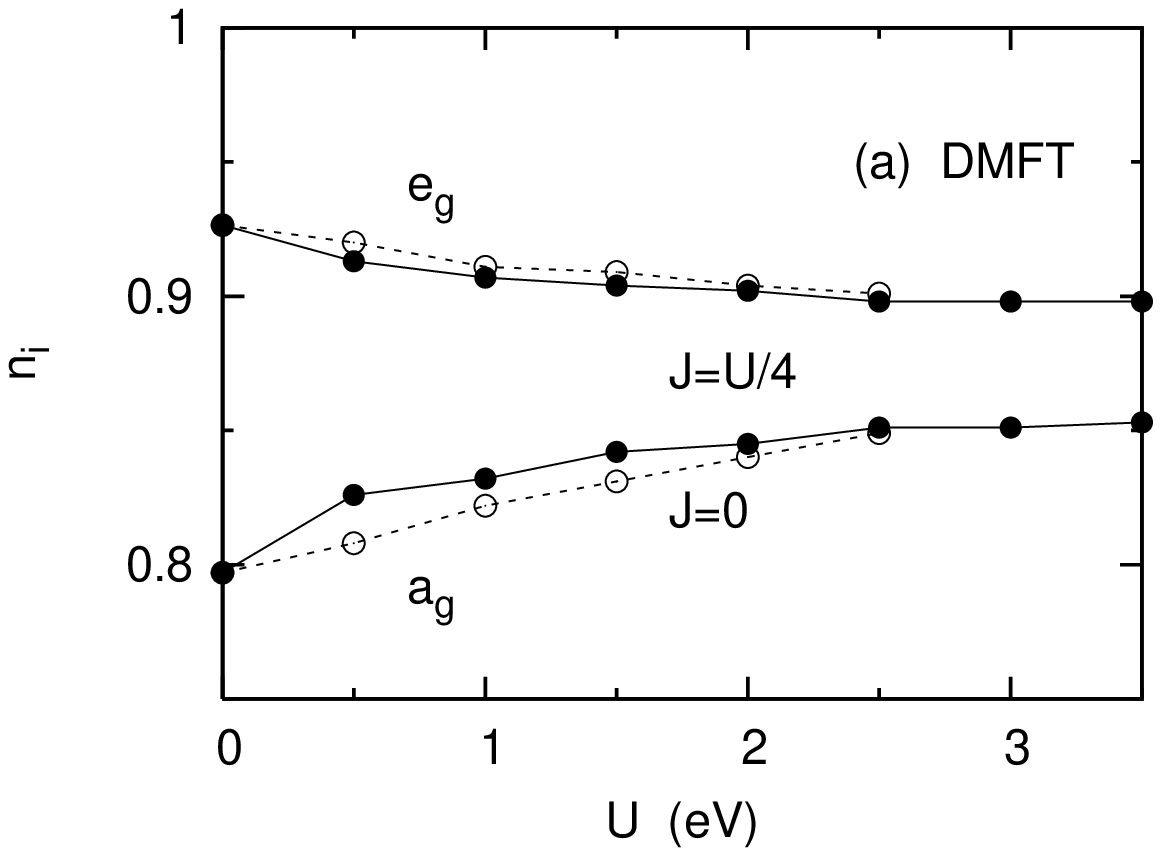}
  \includegraphics[width=4.8cm,height=7cm,angle=-90]{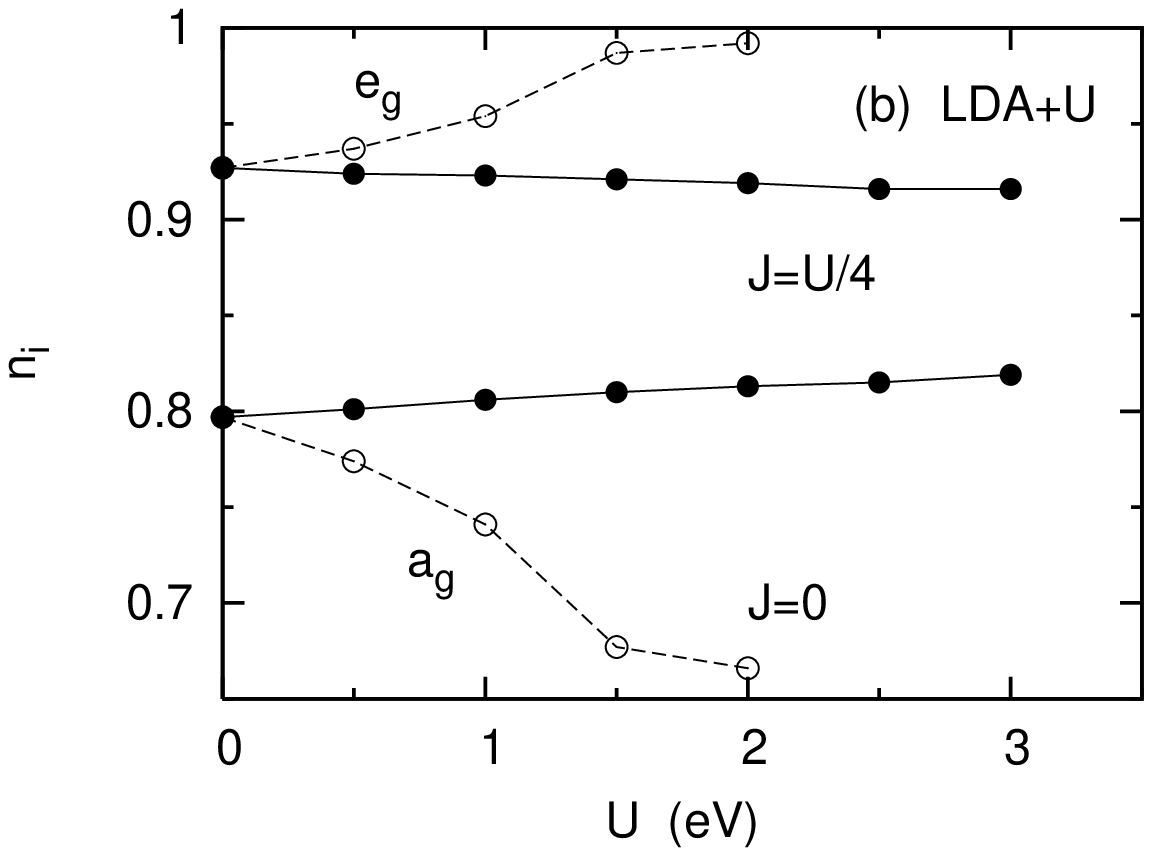}
  \end{center}
  \vskip-4mm
\caption{
Occupations of $a_{1g}$ and $e_{g'}$ subbands of Na$_{0.3}$CoO$_2$ as a
function of $U$. (a) DMFT; (b) LDA+U. Solid dots: $J=U/4$, empty dots:
$J=0$. The lines are guides for the eye. 
}\end{figure}

To illustrate the variation of the $t_{2g}$ subband occupations with local 
Coulomb and exchange energies, we show in Fig.~3(a) the trend obtained within 
the DMFT for $U'=U/2$, $J=U/4$ and $U'=U$, $J=0$. In both cases 
charge transfer proceeds from $e_{g'}$ to $a_{1g}$, i.e., orbital     
polarization is reduced. We have evaluated the DMFT quasi-particle 
spectra both in the non-diagonal $t_{2g}$ and diagonal $a_{1g}$, $e_{g'}$ 
representations. Both versions are in excellent agreement, indicating that
in the present system non-diagonal coupling among  $t_{2g}$ states is fully
taken into account within the $a_{1g}$, $e_{g'}$ representation. 

The variation of the $a_{1g}$, $e_{g'}$ subband occupations with $U$ derived
within the DMFT differs from the one found in the LDA+U, 
as illustrated in Fig.~3(b). While dynamical correlations lead to reduced 
orbital polarization for $J=U/4$ and $J=0$, the LDA+U treatment gives this
trend only for $J=U/4$; $J=0$ yields the opposite effect.  
This dependence of the orbital polarization on the ratio 
$J/U$ within the LDA+U follows from the Hartree Fock self-energy. 
For a paramagnetic $t_{2g}$ complex with one-fold $a_{1g}$ 
and two-fold $e_{g'}$ subbands, the orbital dependent potential is given by 
\cite{lie}:
\,$V^{\rm LDA+U}_{i\ne j} = \Sigma_{ij}^{\rm HF} = \delta (U-5J)$, 
with \,$\delta = (n_e-n_a)/3$. The diagonal term 
\,$\Sigma_{ii}^{\rm HF} = 5\bar n (U-2J)$\, 
gives an overall energy shift, where $\bar n = (n_a + 2 n_e)/3$.
Within the $a_{1g}$, $e_{g'}$ basis, the self-energies are
\,$\Sigma_{a}^{\rm HF} = \Sigma_{ii}^{\rm HF} + 2\Sigma_{ij}^{\rm HF}$, 
\,$\Sigma_{e}^{\rm HF} = \Sigma_{ii}^{\rm HF} -\Sigma_{ij}^{\rm HF}$.
Subtracting the diagonal term the shifted band energies are:
\ $\epsilon'_a(k) = \epsilon_a(k) + 2 \delta (U-5J)$ and
\ $\epsilon'_e(k) = \epsilon_e(k) -   \delta (U-5J)$.
Thus, for $n_e>n_a$ the $a_{1g}$ ($e_{g'}$) energies are shifted up 
(down) as long as $J<U/5$. For $J>U/5$ this trend is reversed. For
realistic $U=3.0\cdots3.5$~eV, $J\approx0.8$~eV, the latter condition
is satisfied, implying reduced orbital polarization also in the static
limit.   

Comparing the static self-energy with the one obtained in second-order
perturbation theory \cite{lie} we find that, because of the small 
difference between $n_a$ and $n_e$, for increasing $U$ static 
correlations are rapidly dominated by dynamical correlations.
Also, the more compact $e_{g'}$ density of states ensures that 
${\rm Re}\,\Sigma^{(2)}_e(\omega=0)>{\rm Re}\,\Sigma^{(2)}_a(\omega=0)$
for $J=0$ and $J=U/4$, implying diminishing orbital polarization 
\cite{ogata}. Thus, static and dynamical correlations exhibit 
qualitatively different dependencies on the ratio $J/U$. In addition, 
the reduced orbital polarizations found for $J=U/4$ in the static 
and dynamical cases arise for different physical reasons.

\begin{figure}[t!]
  \begin{center}
  \includegraphics[width=4.8cm,height=7cm,angle=-90]{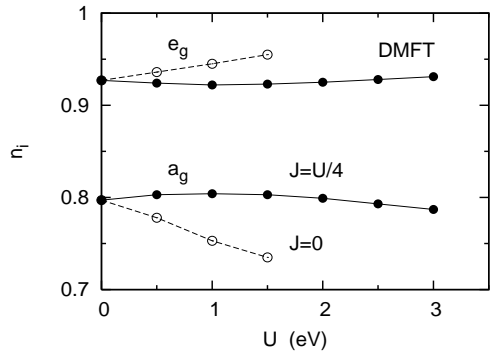}
  \end{center}
  \vskip-4mm
\caption{
Occupations of 3-band model: elliptical density of states; band fillings 
as in Na$_{0.3}$CoO$_2$; DMFT. Solid dots: $J=U/4$, empty dots: $J=0$. 
The lines are guides for the eye. 
}\end{figure}

The results in Fig.~3 demonstrate that in the metallic phase it is crucial
to include dynamical correlations. Evidently, the new degrees of freedom
generated by quasi-particle interactions, such as spectral weight transfer
between low and high frequencies, relaxation shifts (band narrowing) and
decay processes, which are beyond the LDA and LDA+U treatments, contribute
to the charge balance between non-equivalent orbitals. It is to be expected 
that these dynamical correlations also depend on the single-particle bands,
i.e., on the available density of occupied and unoccupied states involved in
excitation processes. To illustrate this point we show in Fig.~4 the  
subband occupations for an analogous 3-band model with elliptical density 
of states and total occupancy 5.3. Adjusting the chemical shift to simulate 
the $a_{1g}$, $e_{g'}$ occupations of  Na$_{0.3}$CoO$_2$ we find
increasing orbital polarization for $J=0$ but weakly varying $n_i$ for 
$J=U/4$ \cite{comment}. 
Clearly, the inter-orbital charge transfer depends on the 
combination of single- and many-particle interactions. 

The correlation induced band narrowing seen in Fig.~2 is consistent 
with infrared optical data \cite{wang} which show transitions within 
the $t_{2g}$ manifold at lower energies than predicted by the LDA 
\cite{johannes}. Substantial band narrowing is also observed in all 
ARPES measurements \cite{hasan1,yang1,yang2,hasan2}. On the other hand,
the present results are in conflict with the ARPES data as far as the 
filling of the $e_{g'}$ bands is concerned. So far, we have no explanantion 
for this qualitative discrepancy between our state-of-the-art
DMFT results and the ARPES data. The interpretation
of photoemission spectra can be complicated due to surface induced 
single-particle and correlation features \cite{surface}. Moreover, 
matrix element effects could play a role. For instance, along $\rm\Gamma K$ 
(chosen as $x$ axis) the $a_{1g}$ and $e_{g2}$ bands are even with respect 
to $\pm(y,z)$ where $z$ defines the surface normal (see Fig.~1). In contrast, 
$e_{g1}$ is odd. These states therefore couple differently to the incident 
photon field. To identify the orbital character of the $t_{2g}$ bands near 
$E_F$ the use of various polarizations is recommended.    

It is interesting to compare the influence of dynamical correlations on the 
subband occupations of Na$_{0.3}$CoO$_2$ to those found in other multi-band 
materials. The $t_{2g}$ 
valence bands of the layer perovskite Sr$_2$RuO$_4$ also have $a_{1g}$, 
$e_{g'}$ symmetry, but are $2/3$ filled. Because of the smaller width of the 
$e_{g'}$ bands DMFT predicts a charge transfer to the $a_{1g}$ band \cite{lie}, 
shifting the $a_{1g}$ van Hove singularity closer to $E_F$ than in the LDA 
\cite{mazin}. The basic shape of the Fermi surface is preserved. The 
transfer from $e_{g'}$ to $a_{1g}$ agrees with the one found here for 
Na$_{0.3}$CoO$_2$. A similar inter-orbital charge transfer was recently
obtained in DMFT calculations for BaVS$_3$ \cite{lechermann}. In the 
orthorhombic phase this $3d^1$ material exhibits a wide $a_{1g}$ band and
weakly occupied, narrow $e_{g'}$ bands. Dynamical correlations were found to 
cause a transfer of electrons from $a_{1g}$ to  $e_{g'}$ states, thereby 
reducing orbital polarization. As in the present system, the same trend was 
obtained for small and large Hund's rule exchange terms ($J=U/7,\ U/4$).
On the other hand, as recently shown by Pavarini {\it et al.} \cite{pavarini} 
for several $3d^1$ perovskites, non-diagonal $t_{2g}$ coupling caused by 
octahedral distortions decreases orbital fluctuations and enhances insulating 
behavior. Also, Manini {\it et al.}~\cite{manini} studied the Mott transition 
in a model consisting of two equal subbands with unit total occupancy. 
With increasing chemical shift, causing more diverse band fillings, 
the tendency towards a Mott insulator was found to be enhanced. 
Since in their case $J=0$, these results are consistent with the enhanced
orbital polarization shown in Fig.~4 for $J=0$. These various trends underscore 
the subtle nature of correlations in multi-band systems, and their remarkable 
sensitivity to various system parameters, in particular, the shape of the 
density of states and the size of $J/U$.   

In summary, we have explored the modification of the Fermi surface of 
Na$_{0.3}$CoO$_2$ as a result of dynamical correlations.
The main effect of these fluctuations
is the spectral weight transfer from the quasi-particle peak near 
$E_F$ to the incoherent part of the spectrum.
In a multi-band material, these dynamical processes 
are orbital dependent. In particular, since they depend on orbital 
occupations, they can give rise to charge transfer between subbands
induced via hybridization and interband Coulomb interactions. The 
highlight of the LDA band structure of Na$_{0.3}$CoO$_2$ is the 
distinctly different filling of the $t_{2g}$ bands, with the nearly
full $e_{g'}$ subbands yielding characteristic hole pockets
of the Fermi surface. Accounting for correlations within the 
DMFT we have shown that electronic charge 
is shifted from the $e_{g'}$ subbands to the $a_{1g}$ band, thus
slightly enlarging the small $e_{g'}$ hole pockets and reducing the 
main $a_{1g}$ pocket centered at $\rm \Gamma$. Further studies are 
needed to reconcile these theoretical findings with the photoemission 
data.

We like to thank O. Gunnarsson and I. I. Mazin for useful comments. 

Email address: \hfill 
ishida@chs.nihon-u.ac.jp; 
johannes@ dave.nrl.navy.mil;
a.liebsch@fz-juelich.de

\end{document}